\documentclass[a4paper]{jpconf}
\usepackage{graphicx}
\begin{document}
\title{Possible indicators for low dimensional superconductivity in the quasi-1D carbide Sc$_3$CoC$_4$ }

\author{E--W Scheidt, C Hauf, F Reiner, G Eickerling, and W Scherer}

\address{CPM, Institut f\"{u}r Physik, Universit\"{a}t Augsburg, 86159
Augsburg, Germany}

\ead{Ernst-Wilhelm.Scheidt@physik.uni-augsburg.de}

\begin{abstract}
The transition metal carbide Sc$_3$CoC$_4$ consists of a
quasi-one-dimensional (1D) structure with  $[$CoC$_4]_{\infty}$
polyanionic chains embedded in a scandium matrix. At ambient temperatures
Sc$_3$CoC$_4$ displays metallic behavior. At lower temperatures, however,
charge density wave formation has been observed around 143\,K which is
followed by a structural phase transition at 72\,K. Below $T_{\rm{c}}^{onset}= 4.5$\,K the
polycrystalline sample becomes superconductive. From $H_{\rm{c1}}(0)$
and $H_{\rm{c2}}(0)$ values we could estimate the London penetration depth
($\lambda_{\rm{L}} \cong 9750$\,\AA) and the Ginsburg-Landau (GL)
coherence length ($\xi_{\rm{GL}} \cong 187$\,\AA). The resulting
GL-parameter ($\kappa \cong 52$) classifies Sc$_3$CoC$_4$ as a
type II superconductor. Here we compare the puzzling superconducting
features of Sc$_3$CoC$_4$, such as the unusual temperature dependence
i) of the specific heat anomaly and ii) of the upper critical field $H_{\rm{c2}}(T)$ at $T_{\rm{c}}$,  and iii) the
magnetic hysteresis curve, with various related low dimensional
superconductors: e.g., the quasi-1D superconductor  (SN)$_x$ or
the 2D transition-metal dichalcogenides. Our results identify
Sc$_3$CoC$_4$ as a new candidate for a quasi-1D superconductor.
\end{abstract}

\section{Introduction}

Low dimensional superconductivity is a vital and controversial research
topic in solid sate physics and chemistry. However, beside the well-established
examples for quasi-one-dimensional (1D) superconductors, e.\,g.,
polysulfur nitride (SN)$_x$ \cite{Lou1989},
the organic Bechgaard salts \cite{Jerome1982} or the transition-metal
trichalcogenides \cite{Srivastava1992}, there are only a few
systems known which allow us to study the nature of the phenomenon in
greater detail. Recently, we observed superconductivity
in Sc$_3$CoC$_4$ which might represent a new benchmark system of a quasi-1D superconductor
\cite{Scherer2010}. Sc$_3$CoC$_4$ consists of quasi-one-dimensional
$[$CoC$_4]_{\infty}$ polyanionic chains embedded in a scandium
matrix \cite{Rohrmoser2007}. From specific heat, magnetic susceptibility,
resistivity and X-ray diffraction measurements a charge density wave around 143\,K
and a structural phase transition below 72\,K were identified
\cite{Scherer2010}. Therefore, it is very natural to ask if the
superconductivity of this quasi-1D carbide is also of low dimensional character.
In the absence of a single crystal, we compare in this paper i) structural
features and physical properties like ii) the specific heat anomaly of the
superconducting transition, iii) the magnetization behavior, and iv) the
temperature dependence of the upper critical field with the established low
dimensional superconductors like polysulfur nitride (SN)$_x$, the Bechgaard salts, high-$T_{\rm{c}}$-cuprates,
and the dichalcogenide NbSe$_2$ \cite{Soto2007}.

\section{Structural details}

\begin{figure}[h]
\begin{center}
\includegraphics[width=130mm]{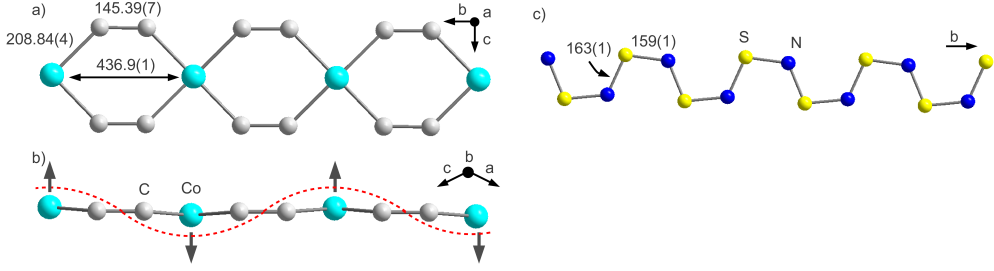}
\caption{\label{fig1} The $[$CoC$_4]_{\infty}$\,ribbons of
Sc$_3$CoC$_4$ a) in the crystallographic \emph{b},\emph{c} plane
and b) in the low temperature modification (side view) \cite{Scherer2010}.
c) The $[$SN$]_{\infty}$\,ribbons of the polysulfur nitride (SN)$_x$
showing a zigzag pattern with alternating short and long S-N distances
\cite{Marshall1976}. Selected bond distances are given in pm.}
\end{center}
\end{figure}
High resolution single crystal X-ray diffraction studies at room
temperature provide precise structural parameters of Sc$_3$CoC$_4$
(space group $I$mmm) \cite{Rohrmoser2007}: the Co atoms are coordinated
by four C$_2$ pairs in an almost square planar manner, establishing
one-dimensional infinite $[$CoC$_4]_{\infty}$ chains as
pictured in Fig.\,\ref{fig1}a. Below the structural phase transition
at 72\,K Sc$_3$CoC$_4$ is characterized by an alternating displacement
of the Co atoms above and below the $[$CoC$_4]_{\infty}$\,
ribbons (Fig.\,\ref{fig1}b, space group $C$\,2/m). Such a zigzag chain
deformation pattern is also observed in quasi-1D superconductors like
polysulfur nitride (SN)$_x$ (Fig.\,\ref{fig1}c)\cite{Marshall1976},
or the organic Bechgaard salts \cite{Thorup1981}. Therefore, the zigzag
distortion of the $[$CoC$_4]_{\infty}$\,ribbons might be a
structural prerequisite of the low-dimensional superconductivity,
even though the zigzag pattern in Sc$_3$CoC$_4$ is only weakly pronounced in
comparison to (SN)$_x$. However, for the non superconducting analogues
Sc$_3$NiC$_4$ and Sc$_3$FeC$_4$ no structural phase transition was observed above 2\,K
\cite{Scherer2010}.

\section{Experimental results and discussion}
\begin{figure}[h]
\begin{center}
\includegraphics[width=65mm]{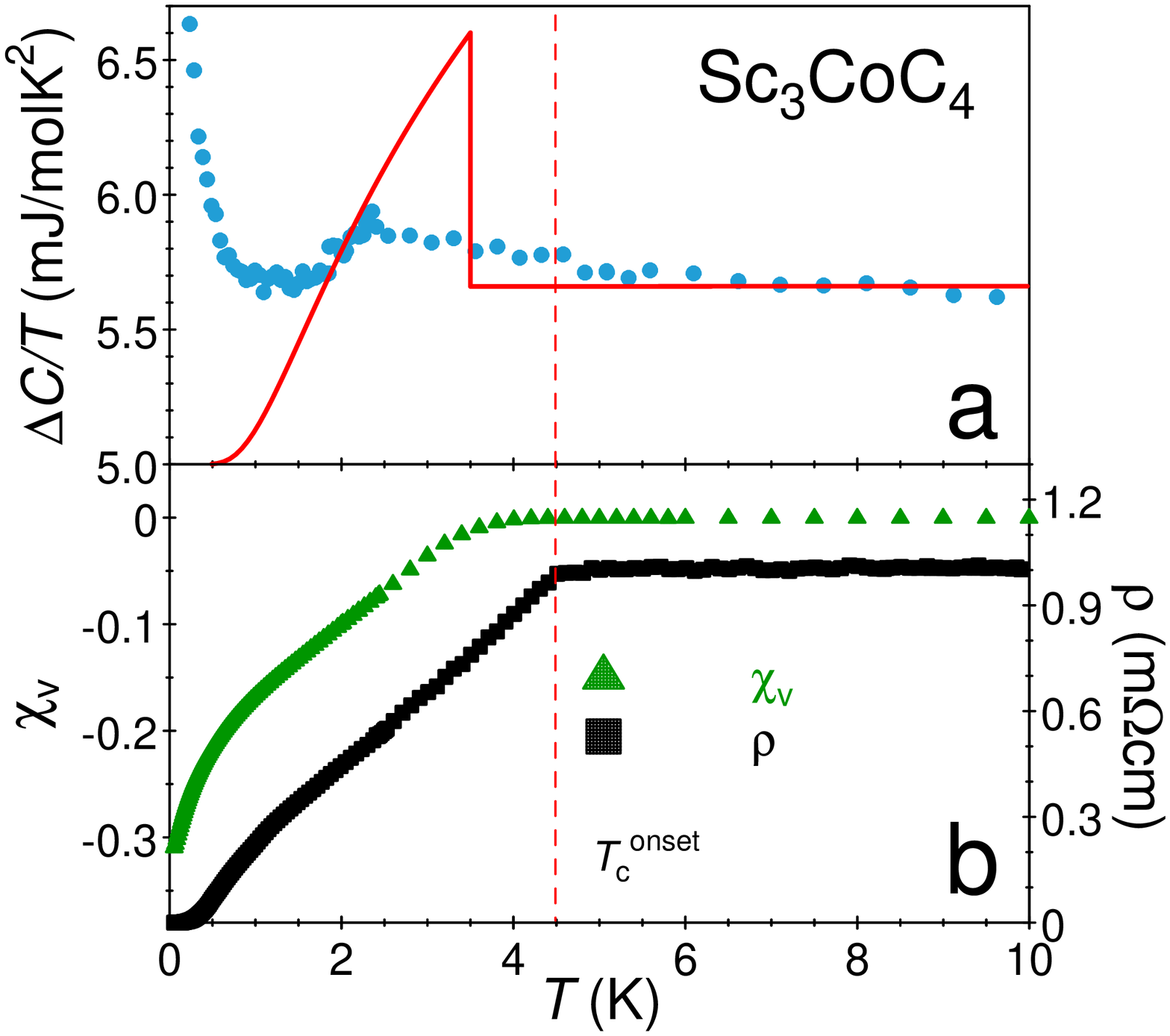}%
\hspace{5mm}%
\includegraphics[width=65mm]{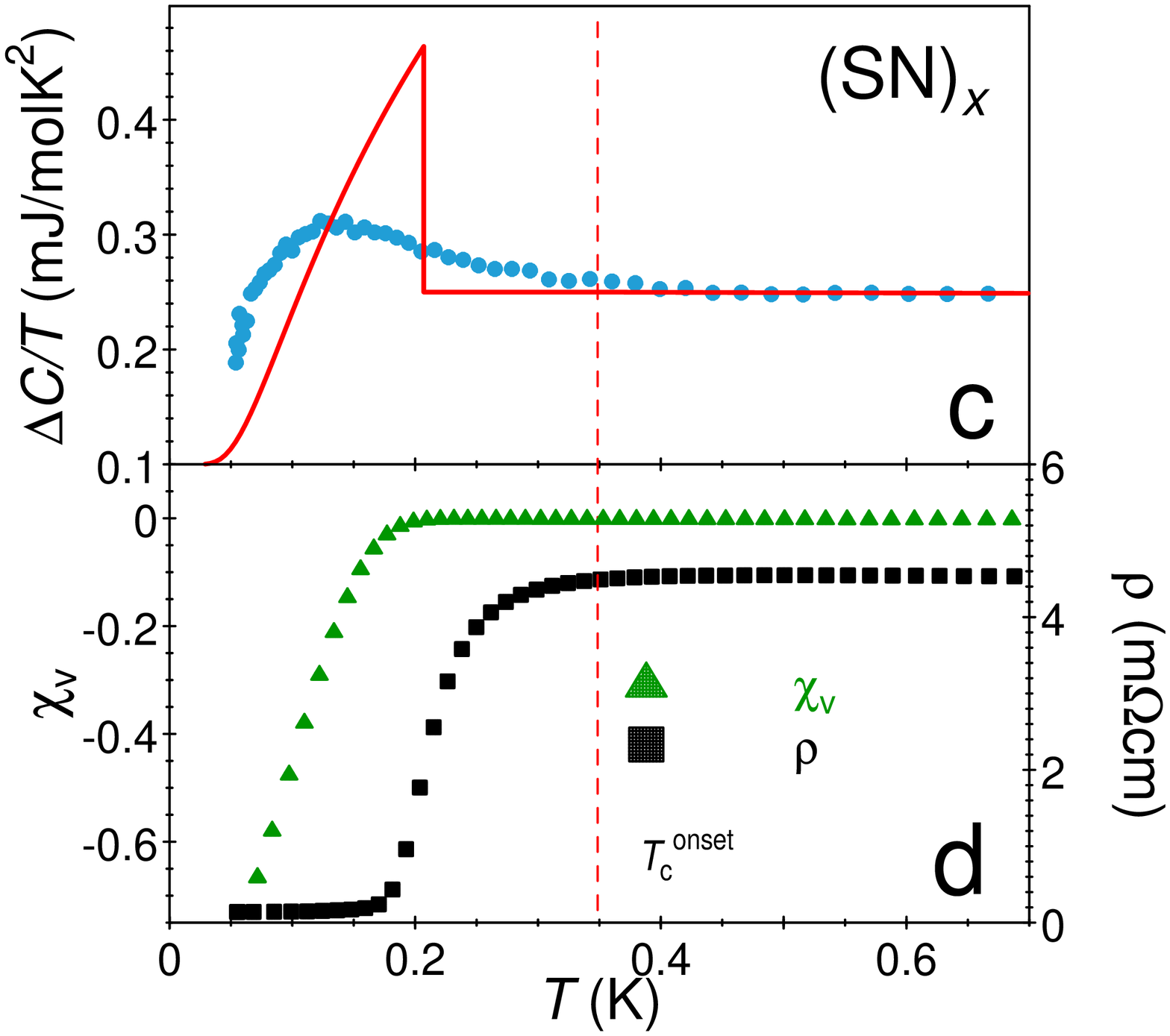}
\caption{\label{fig2}Temperature dependence of a) the electronic contribution
of the specific heat divided by temperature, $\Delta C/T$, and b) the volume
susceptibility, $\chi_{\rm{V}}$, at $H = 1$\,Oe (left scale; triangles) and the electrical
resistivity, $\rho$, (right scale; squares) for Sc$_3$CoC$_4$ in comparison with the
temperature dependence of the same physical properties, c) $\Delta C/T$ and
d) $\chi_{\rm{V}}$, $\rho$ for the quasi-1D superconductor (SN)$_x$. The solid lines
represent the superconducting transition as it is predicted by the BCS model \cite{Muehlschlegel1959}.}
\end{center}
\end{figure}
The low temperature electronic specific heat of Sc$_3$CoC$_4$, shown as
$\Delta C/T$\,vs.\,$T$ plot in Fig.\,\ref{fig2}a, reveals a broad
superconducting transition below $T_{\rm{c}}^{onset}= 4.5$\,K which
was determined by resistivity measurements (Fig.\,\ref{fig2}b).
Additionally, this transition is indicated by the magnetic susceptibility
showing a lowering of the magnetic response just below 4.4\,K. These
results and the fact, that the volume susceptibility at 50\,mK
approaches -0.3, identifies Sc$_3$CoC$_4$ as a bulk superconductor
(Fig.\,\ref{fig2}b). The steep increase of $\Delta C/T$ below 1\,K
is mainly due to hyperfine interactions caused by the nuclear quadrupole
moments of the Sc and Co atoms.

For the superconducting transition in Sc$_3$CoC$_4$ we calculated the expected ideal BCS curve
of a weak coupled superconductor, taking into account the entropy balance between
the normal and superconducting state (solid line in Fig.\,\ref{fig2}a)
\cite{Muehlschlegel1959}. This calculation indicates, that only
a small part ($\Delta \gamma_{\rm{n}} = 0.65\pm0.05$\,mJ/mol\,K$^2$) of the
total electronic specific heat $\gamma_{\rm{n}} = 5.7$\,mJ/molK$^2$,
which was graphically determined from a $C/T$ vs. $T^2$ plot, can be
involved in the development of the superconducting state. Therefore,
the bulk superconductivity may be related solely to small areas
of the Fermi-surface. Furthermore, the overall shape of the specific heat anomaly
differs substantially from the BCS behavior, missing the characteristic pronounced
jump at $T_{\rm{c}}$.

The unusual broad shape of this anomaly and the fact, that
only a small fraction (12\,$\%$ in Sc$_3$CoC$_4$) of the band electrons contribute to the superconducting
state is also observed in the well known quasi-1D superconductor polysulfur
nitride (SN)$_x$ (Fig.\,\ref{fig2}c). Our specific heat measurement
on a single crystal (SN)$_x$  sample indicates that 60\,$\%$
of the total electronic specific heat ($\gamma_{\rm{n}} = 0.25$\,mJ/molK$^2$)
is involved in the superconducting pairing process. The volume susceptibility at 70\,mK
approaches only a value of -0.66 of the full diamagnetic response (Fig.\,\ref{fig2}d).
These analogies suggest that the unusual broad shape
of the specific heat anomaly and the small percentage of band electrons contributing to the
superconducting state may be characteristic features
for the presence of quasi-one-dimensional superconductivity in Sc$_3$CoC$_4$.

\begin{figure}[htb]
\begin{center}
\includegraphics[width=55mm]{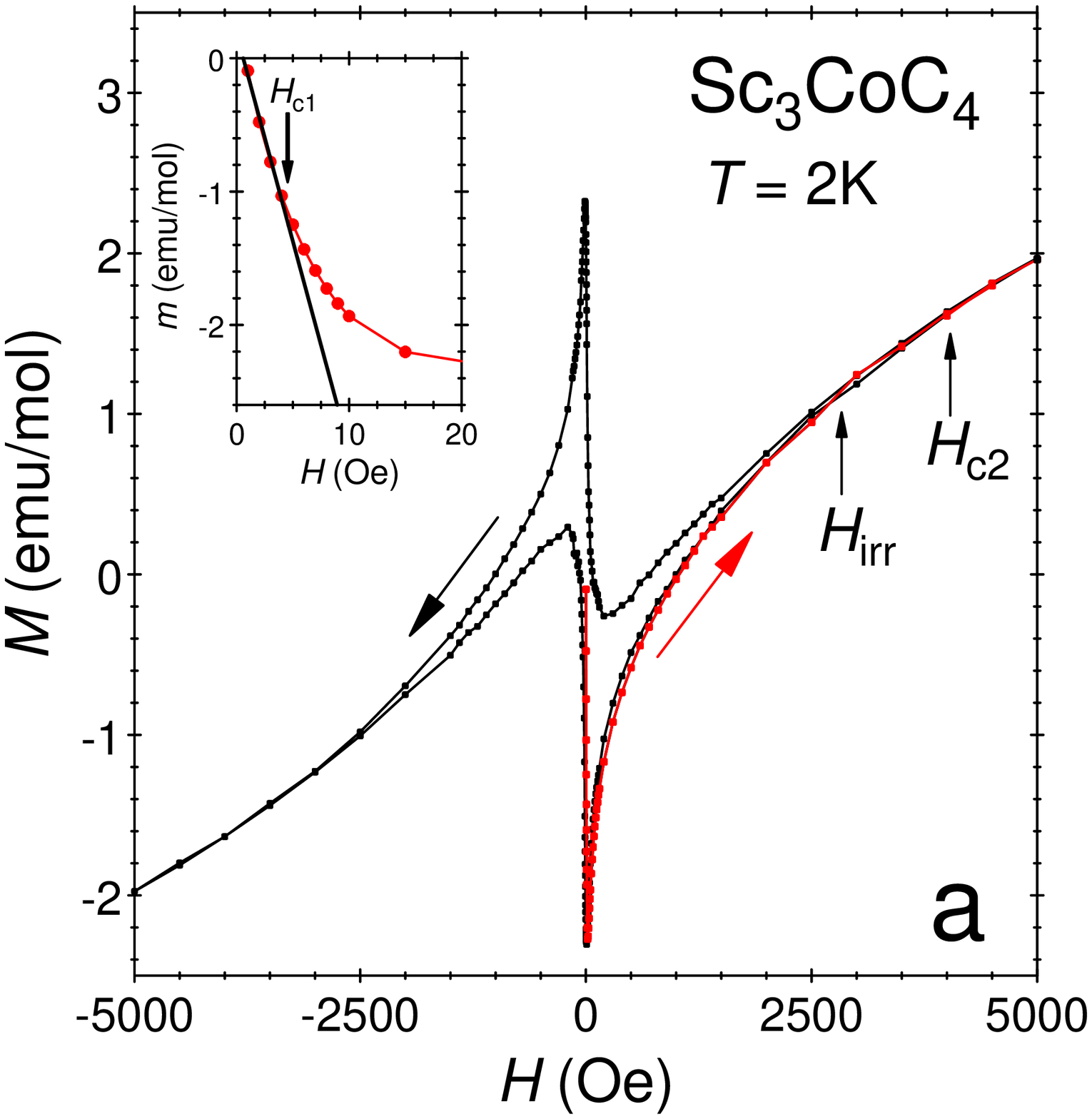}%
\hspace{5mm}%
\includegraphics[width=56.5mm]{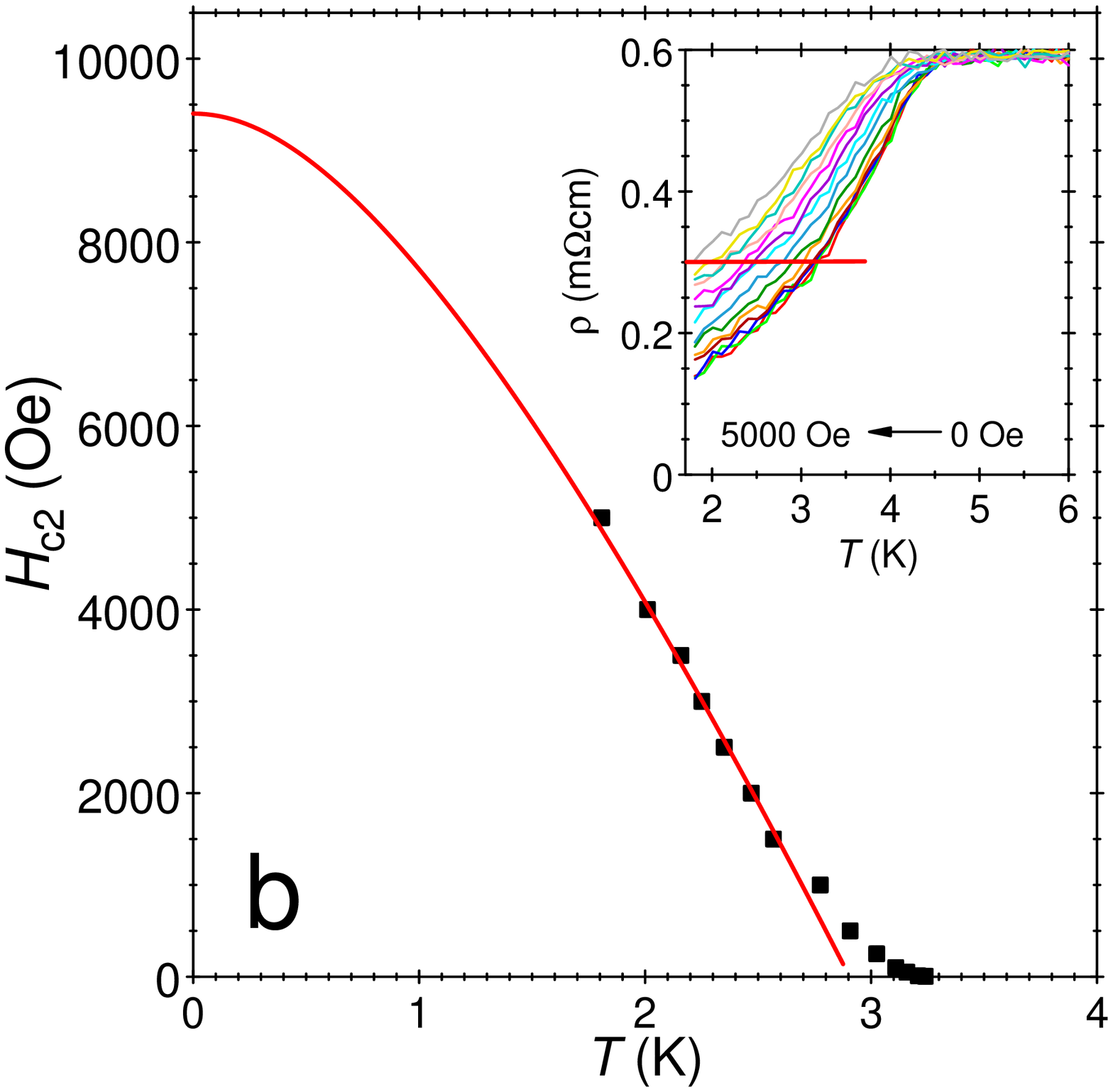}
\caption{\label{fig3}a) Magnetic hysteresis loop of Sc$_3$CoC$_4$ at 2\,K. The
underlying pronounced slope ($M/H$) and the deviation from the
linear behavior of the magnetization is possibly due to
the Pauli susceptibility of the normal conducting electrons and of a small
impurity concentration of elementary Co, respectively. To determine
$H_{\rm{c1}}$, the insert displays an expanded region of the initial
magnetization curve up to 20 Oe. b) Phase diagram of the upper critical
fields $H_{\rm{c2}}(T)$. The solid line corresponds to a WHH-fit
\cite{Werthamer1966}. $T_{\rm{c}}(H_{\rm{c2}})$ is obtained from the  50\,$\%$
values of the normal state resistivity (solid line in the insert). }
\end{center}
\end{figure}

A further hint for low-dimensional superconductivity in Sc$_3$CoC$_4$ presents
the magnetic hysteresis loop (Fig.\,\ref{fig3}a). This  magnetization curve
clearly indicates a type-II superconducting behavior. The characteristic
dip near zero field crossings and the high reversibility
between the irreversibility field ($H_{\rm{irr}}(2K) = 2800 \pm 400$\,Oe)
and the upper critical field ($H_{\rm{c2}}(2K) = 4000 \pm 200$) is also found in
layered cuprate superconductors, but only at higher temperatures \cite{Senoussi1988}.
In Sc$_3$CoC$_4$ and also in the 2D highly anisotropic dichalcogenide superconductors
like NbSe$_2$ \cite{Soto2007, Sonier1999} such a reversible behavior above a
threshold field occurs at very low temperatures and therefore might not only
be due to granularity \cite{Senoussi1988} but also due to the existence of low-dimensionality.

From $H_{c1} =  4.4 \pm 0.2$\,Oe at 2\,K, as determined by the deviation from
the initial linearity of the magnetization curve (insert in Fig.\,\ref{fig3}a),
the lower critical field $H_{c1}(0)$ is estimated to be $7 \pm 0.4$\,Oe.
$H_{\rm{c2}}(0) = 94000 \pm 500$ is calculated from the
WHH-theory \cite{Werthamer1966}, with a Maki parameter of $\alpha =0.25$ estimated from the slope of the
upper critical field at $T_{\rm{c}}$
and the spin-orbit parameter $\lambda_{\rm{so}}$ assumed to be zero.
By combining the upper and lower critical field values with the Ginsburg Landau (GL)
expressions for isotropic superconductors ($H_{\rm{c2}} = \Phi_{0}/(2\pi \xi_{\rm{GL}}^2)$;
$H_{\rm{c1}} = (\Phi_{0}/(4\pi \lambda_{\rm{L}}^2))(0.08+ln\kappa)$) we obtained
the London penetration depth of $\lambda_{\rm{L}} = 9750 \pm 300$\,\AA\,\,and
the GL-coherence length of $\xi_{\rm{GL}} =187 \pm 5$\,\AA. The resulting GL parameter $\kappa = 52$
classifies Sc$_3$CoC$_4$ as a type-II superconductor.

A further interesting effect is the upturn of $H_{\rm{c2}}(T)$  at
temperatures close to $T_{\rm{c}}$ (Fig.\,\ref{fig3}b).
Beside other explanations, e.g., sample imperfections, this effect
is mainly attributed to a reduction of the dimensionality.
This upturn is also found in the quasi-1D superconductors
(SN)$_x$ \cite{Azevedo1976}, the Bechgaard salts \cite{Lee1997} and as R. A. Klemm summarised:
"seems to be a general feature of many, but not all highly
anisotropic materials" \cite{Klemm1985}.

\section{Summary}
Resistivity, magnetization and specific heat measurements
clearly support the presence of bulk superconductivity in the quasi-one-dimensional
transition metal carbide Sc$_3$CoC$_4$. Several characteristic features might
identify Sc$_3$CoC$_4$ as a promising system in the rare class of quasi-one-dimensional
superconductors: i) the structural zigzag chain, ii) the broad shape
of the specific heat anomaly, iii) the small percentage of band electrons contributing to the
superconductivity  state, iv) the partially reversibility of the magnetic hysteresis loop, and v) the upturn of the
$H_{\rm{c2}}(T)$ at temperatures near $T_{\rm{c}}$. However, future work on single
crystalline samples of Sc$_3$CoC$_4$ is essential to analyze in detail the
electronic anisotropy in both, the normal and the superconducting state.

\section{Acknowledgments}
We acknowledge valuable discussion with K. L\"uders. This work was supported by the Deutsche Forschungsgemeinschaft (SPP1178).
We thank J. Passmore for providing a (SN)$_x$ sample.\\

\section{References}
\providecommand{\newblock}{}

\end{document}